\documentclass[slac_one]{revtex4}
\usepackage{graphicx}
\usepackage{fancyhdr}
\usepackage{wrapfig,rotating}
\begin{document}

\title{High Mass Resonance Searches at CDF} %%
%Paper title goes here


\author{Kenichi Hatakeyama}
\affiliation{Rockefeller University, New York, NY 10065, USA.\\
for the CDF Collaboration}
%
%\author{P. Lucas}
%\affiliation{FNAL, Batavia, IL 60510, USA}

\begin{abstract}
Recent searches for dijet, dielectron, and dimuon
resonances by the CDF Collaboration are presented.
No evidence for a signal is found in any channel,
so 95\% confidence level upper limits are set on the
new particle production.
\end{abstract}

\maketitle

\thispagestyle{fancy}

\section{Introduction}

Many models beyond the standard model (SM) predict
the presence of new heavy particles which can potentially
be observed as narrow resonances in the invariant mass spectra of
high $p_T$ objects.
The Randall-Sundrum (RS) graviton ($G^*$) in the RS model,
$Z'$, $W'$, 
excited quarks ($q^*$) in the quark compositeness model,
axigluon ($A$) in the chiral color model,
coloron ($C$) in the flavor universal coloron model,
color-octet mass-degenerate techni-$\rho$ ($\rho_{T8}$) in the
technicolor model, and the diquark ($D$) in the $E_6$ model are such
examples.
Below, recent searches for narrow resonances in the dijet and dilepton
channels are presented.

\section{Dijet Mass Resonance Search}

\begin{wrapfigure}{r}{0.45\columnwidth}
\centerline{
\includegraphics[width=0.45\columnwidth,clip=]
{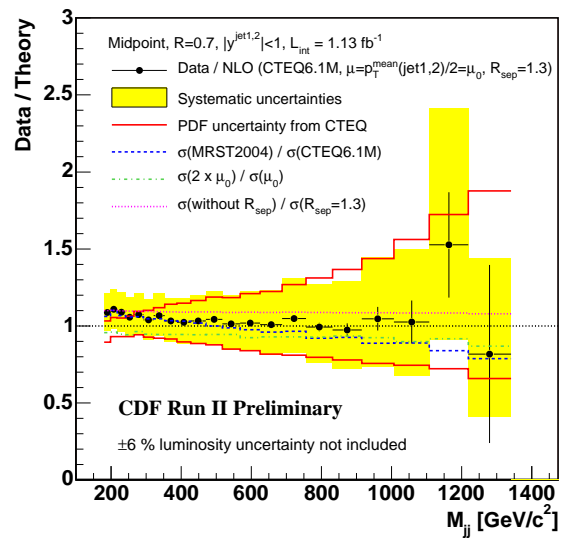}
}
\caption{The ratio of the data to the NLO pQCD predictions versus dijet mass.}
  \label{fig:mjj}
\end{wrapfigure}

CDF recently performed a search for dijet mass resonances using 1.1
$\mbox{fb}^{-1}$ of jet trigger data~\cite{mjj}.
Jets are clustered using the cone-based midpoint jet clustering
algorithm with cone radius $R_{cone} = 0.7$, and we form
dijet mass ($m_{jj}$) spectrum in events with %% containing two leading jets with
$|y^{jet1,2}|<1.0$.
After correcting jet energies for the calorimeter non-linearity
and non-uniformity effects, a $m_{jj}$ spectrum is formed
from the leading two jets, which is
%%This spectrum is 
further corrected for the bin-by-bin migration effect
due to the finite resolution in the $m_{jj}$ measurement.
The background (BG) in this search is dominated by QCD dijet production.
The measured spectrum is compared to
%%and compared to 
the next-to-leading order QCD predictions in Fig.~\ref{fig:mjj}.
They are in good agreement.
%%

%%
%% - BG subtration
%% - Correction
%% - 
%%
Narrow mass resonances are searched for in the measured
dijet mass spectrum by fitting the spectrum to
a smooth functional form 
$d\sigma/dm_{jj} = p_0(1-x)^{p_1}/x^{p_2+p_3\ln(x)}$, 
$x = m_{jj}/\sqrt{s}$,
which provides a good description of dijet mass spectra
from QCD predictions,
and by looking for data points that show significant excess
from the fit.
The fit of this form to the measured spectrum 
shows no significant excess.
% and yields $\chi^2/n.d.f. = 16/17$, so it's concluded that
%the data do not show any evidence for new particle production decaying
%into dijets.

\begin{figure}[htb]\centering\leavevmode
  \includegraphics[width=0.49\hsize,clip=]
  {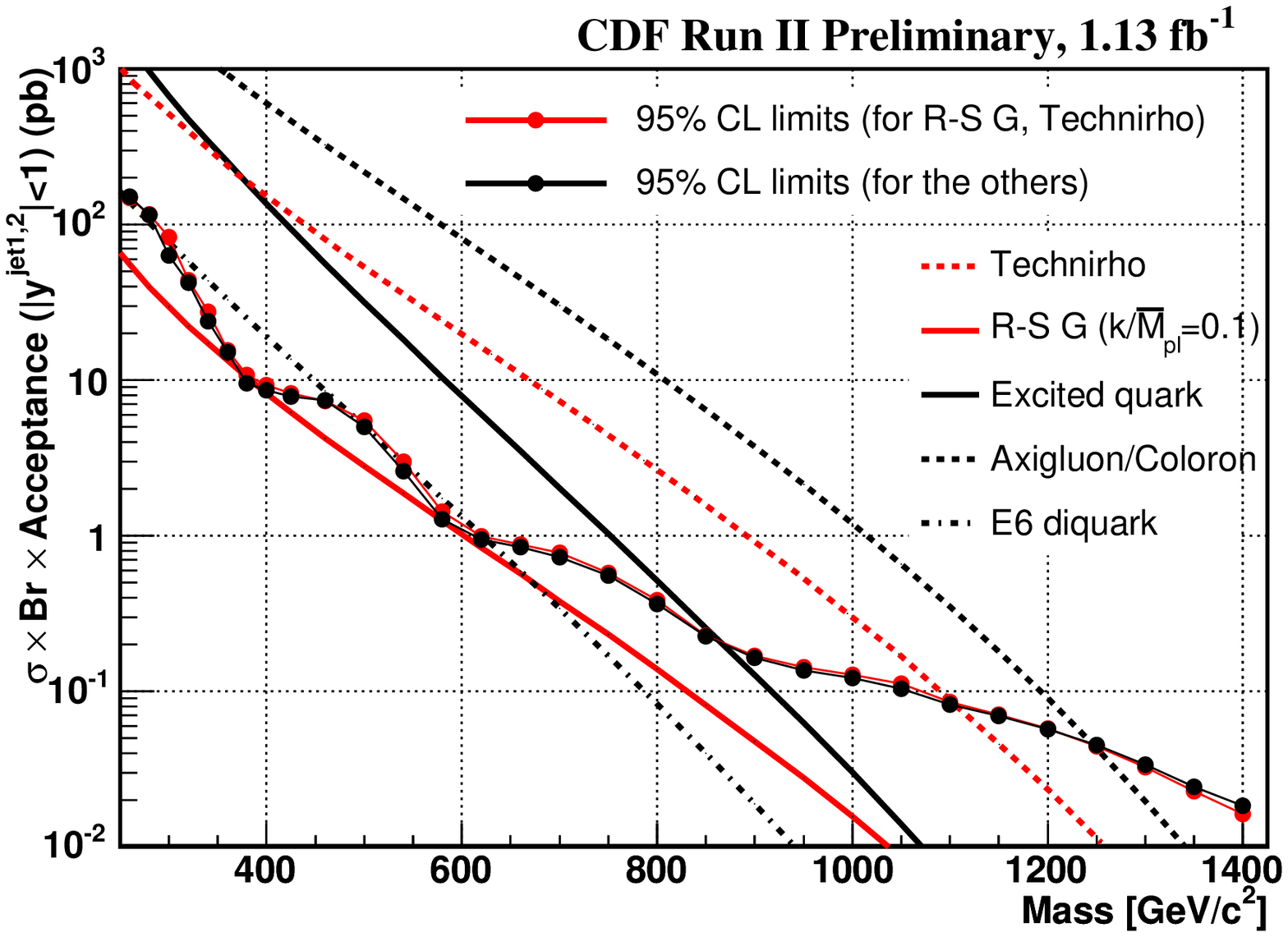}
  \includegraphics[width=0.49\hsize,clip=]
  {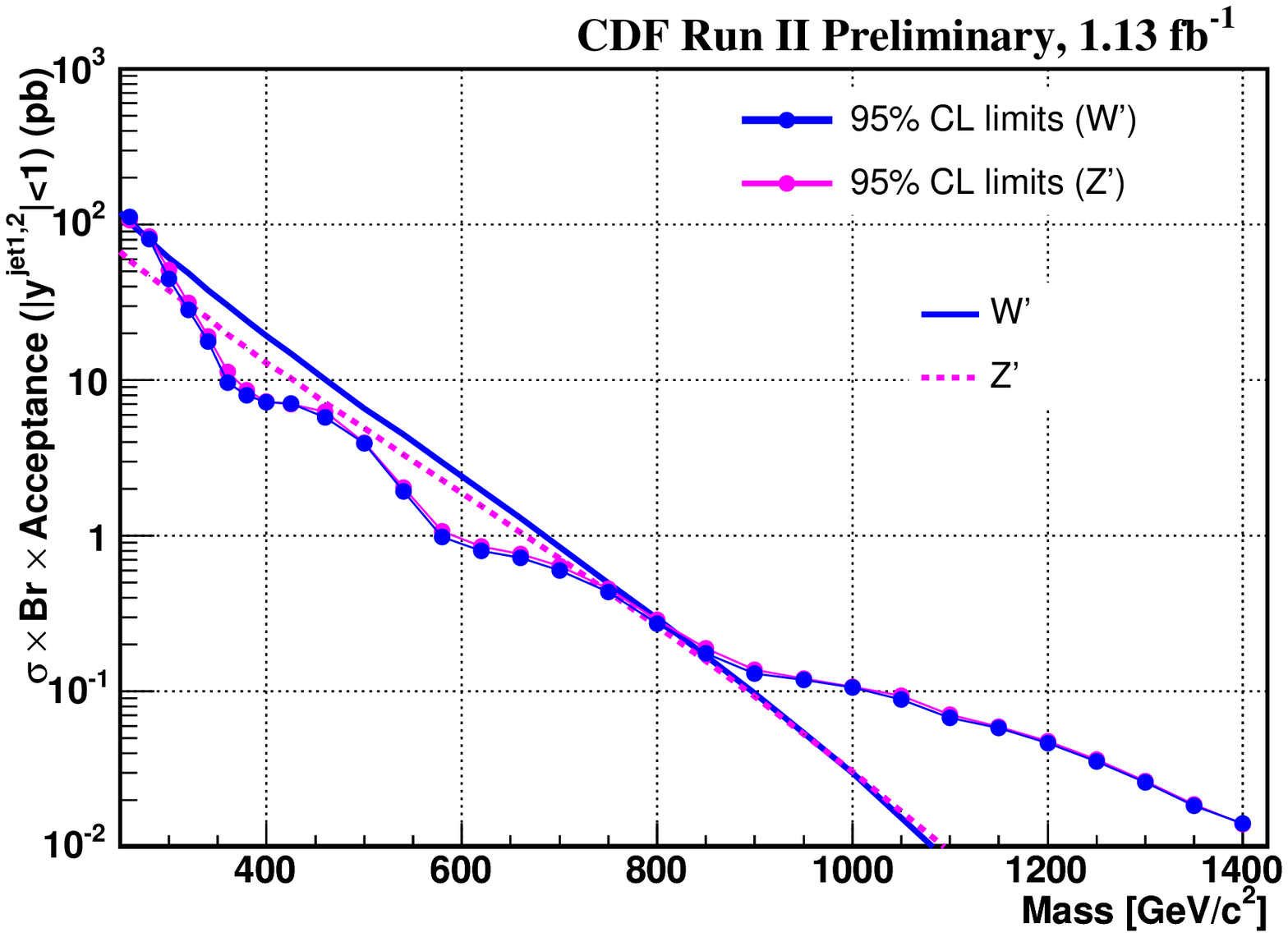}
%   \begin{tabular}{cc}
%   \smash{\lower30.0ex\hbox{\includegraphics[width=0.45\hsize,clip=]
%   {plots/c_mjjVL5_data_NLOJET_ratio_bless.eps}}}&
%   \includegraphics[width=0.39\hsize,clip=]
%   {plots/plot_limit_theory_vs_mass.eps}\\
%   &
%   \includegraphics[width=0.39\hsize,clip=]
%   {plots/plot_limit_theory_vs_mass_wpmzpm.eps}\\
%   \end{tabular}
  \caption{95\% C.L upper limits on $\sigma\cdot{\cal B}(X\to jj)\cdot
    {\cal A}(|y^{jet1,2}|<1)$
 obtained with the signal shapes from the $W'$, $Z'$, $G^*$, 
 and $q^*$ production compared to various theoretical predictions.}
  \label{fig:mjj_limits}
\end{figure}

%
% limits
%
Limits on the new particle production cross sections are set
on $\sigma\cdot {\cal B}(X\to jj) \cdot {\cal
A}(|y^{jet1,2}|<1)$.
The $m_{jj}$ distributions from $q^*$, $G^*$, $W'$, $Z'$ have
different decay modes and thus different signal shapes; therefore,
limits are computed using these four signal shapes separately and
shown in Fig.~\ref{fig:mjj_limits}
together with predictions from various models.
This search sets the most stringent mass exclusions on the production
of $q^*$, $A$, $C$, $D$, and $\rho_{T8}$, which are
%%The mass exclusion regions for all the particles considered are
$260<m_{q^*}<870\mbox{~GeV}/c^2$~\cite{exq_couplings},
$260<m_{A/C}<1250\mbox{~GeV}/c^2$, %for the axigluon and flavor universal
%coloron, %%with the mixing of two $SU(3)$'s, $\cot\theta=1$,
$290<m_{D}<630\mbox{~GeV}/c^2$, and
$260<m_{\rho_{T8}}<1100\mbox{~GeV}/c^2$~\cite{rhot8_para}, respectively.
% for the mass-degenerate
%$\rho_{T8}$
%with the set of parameters discussed earlier,
%$280<m<840\mbox{~GeV}/c^2$ for $W'$, and
%$320<m<740\mbox{~GeV}/c^2$ for $Z'$,
%respectively.
%For the RS gravitons, we could not exclude any mass
%region.

\section{Dielectron Resonance Search}

CDF reported a new search~\cite{mee} for dielectron resonances using
2.5 $\mbox{fb}^{-1}$ of data collected by triggering
on one or two electromagnetic (EM) clusters.
Offline, events are required to have two isolated electrons
with $E_T>25$ GeV,
one in the central ($|\eta|<1.1$) region and the other
one in either the central or the plug ($1.1<|\eta|<2.0$)
regions.
%Events with both electrons in
%the plug regions are not considered in this letter
%since adding them gains little sensitivity.
%
% opposite sign requirement for CC, but not for CP
%
%
%%ffffffffff
%Only electrons with ET greater than 25
%GeV and || < 2 are used in order to ensure 100% trigger
%efficiency and coverage by the the silicon tracker.
%Electrons in the central EM calorimeter are required to
%have a well-measured track in the COT system pointing
%at an energy deposit in the calorimeter. For electrons
%in the plug EM calorimeter, the track association
%uses a calorimeter-seeded silicon-tracking algorithm [8].
%An opposite-charge requirement is applied to electron objects
%pairs detected in the central EM calorimeter. No
%such requirement is applied when one electron is detected
%in the plug, where -dependent charge misidentification
%occurs. The 28 cm radial range of silicon at || >1.1,
%where the COT coverage is incomplete, is insufficient to
%determine accurately the curvature for high pT tracks, as
%predicted by simulation. 

\begin{wrapfigure}{r}{0.49\columnwidth}
\centerline{
\includegraphics[width=0.49\columnwidth,clip=]
{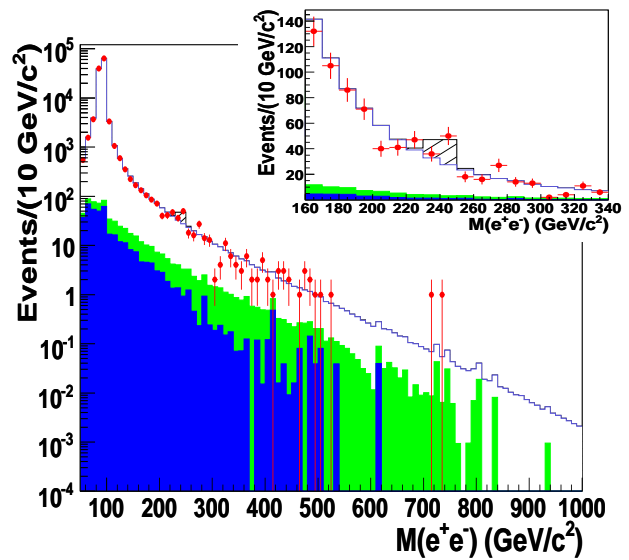}
}
\caption{Observed $e^+e^-$ invariant mass distribution compared to the
  estimated expected backgrounds in a log scale. (inset) The same for
  the 240 GeV/$c^2$ region using a linear scale. The hatched histogram
  shows the expected signal from a 240 GeV/$c^2$ spin-1 particle
  normalized to twice the number of excess events seen in the data.}
  \label{fig:mee}
\vspace*{-2cm}
\end{wrapfigure}

The dominant BG in this search is Drell-Yan production of $e^+e^-$
pairs, which is irreducible. 
Another is dijets and W+jets production (QCD BG)
where one or more jets are misidentified as electrons.
Other contributions include $Z/\gamma^*\to \tau\tau$,
$t\bar t$, and diboson ($W\gamma,WW,WZ,ZZ,\gamma\gamma$) production (other SM BGs).
The Drell-Yan and other SM backgrounds are estimated based on
Monte Carlo (MC),
while the QCD BG is estimated by a data-driven method.

Fig.~\ref{fig:mee} shows the observed $e^+e^-$ invariant mass spectrum from 
2.5 $\mbox{fb}^{-1}$ data together with the expected backgrounds.
The most significant deviation between the data and the SM prediction
occurs at $m_{ee}=241.3$ GeV/$c^2$.
The probability of finding an excess larger than the observed one
from BG only is found to be 0.6\% corresponding to $2.5\sigma$.
The 95\% C.L. limits on $\sigma\cdot{\cal B}(X\to e^+e^-)$
for spin-1 ({\it e.g.}, $Z'$'s) and spin-2 ({\it e.g.}, $G^*$) 
particles obtained from this search are presented in Ref.~\cite{mee}.

%%
%%
%%
% \begin{figure}[tp]\centering\leavevmode
%   \begin{tabular}{cc}
%   \includegraphics[width=0.45\hsize,clip=]
%   {plots/xsec_spin1_paper_fin.eps}&
%   \includegraphics[width=0.45\hsize,clip=]
%   {plots/xsec_spin2_paper_fin.eps}
%   \end{tabular}
%   \caption{$Z+b$-jets differential cross sections as a function of jet
%     $p_T$ (left) and $Z$ $p_T$ (right) from the CDF 2~$\mbox{fb}^{-1}$
%     of data.}
%   \label{fig:ee_limits}
% \end{figure}

\section{Dimuon Resonance Search}

CDF performed a new search for dimuon resonances using 2.3
$\mbox{fb}^{-1}$ of data.
The data are collected online by requiring a track matched with
muon detector hits, and offline, events with two muons with $p_T>30$
GeV/$c$ are selected.

\begin{wrapfigure}{r}{0.49\columnwidth}
\centerline{
\includegraphics[width=0.49\columnwidth,clip=]
{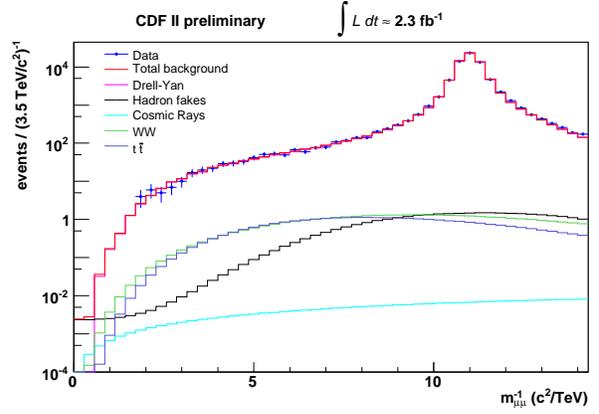}
}
\caption{Observed $1/m_{\mu\mu}$ distribution from 2.3
  $\mbox{fb}^{-1}$ of data compared to the estimated backgrounds.}
  \label{fig:mumu}
\end{wrapfigure}

The main BG in this search comes from Drell-Yan production of dimuons.
The $W^+W^-$ and $t\bar t$ processes also have small contributions.
These contributions are estimated based on Pythia MC.
The hadron fake contributions ($\pi/K$ decays-in-flight) are evaluated
based on same sign dimuon pairs, and
the backgrounds come from cosmic-rays are estimated using track timing
information.

In this search, 
$1/m_{\mu\mu}$ distribution is formed instead of $m_{\mu\mu}$ distribution
as shown in Fig.~\ref{fig:mumu}, since
at high mass, the $m_{\mu\mu}$ resolution is dominated by the track
curvature resolution, resulting in an approximately constant
resolution in $\delta(1/m_{\mu\mu})$.
The signal is searched for by constructing the
$1/m_{\mu\mu}$ distribution templates for $Z'$ boson pole
masses, 
adding the background distributions to the templates, 
and comparing them to the $1/m_{\mu\mu}$ distribution from data in the
search region $1/m_{\mu\mu} < 10$ (TeV/$c^2$)$^{-1}$.

\begin{figure}[hb]\centering\leavevmode
  \begin{tabular}{c}
  \includegraphics[width=0.49\hsize]
  {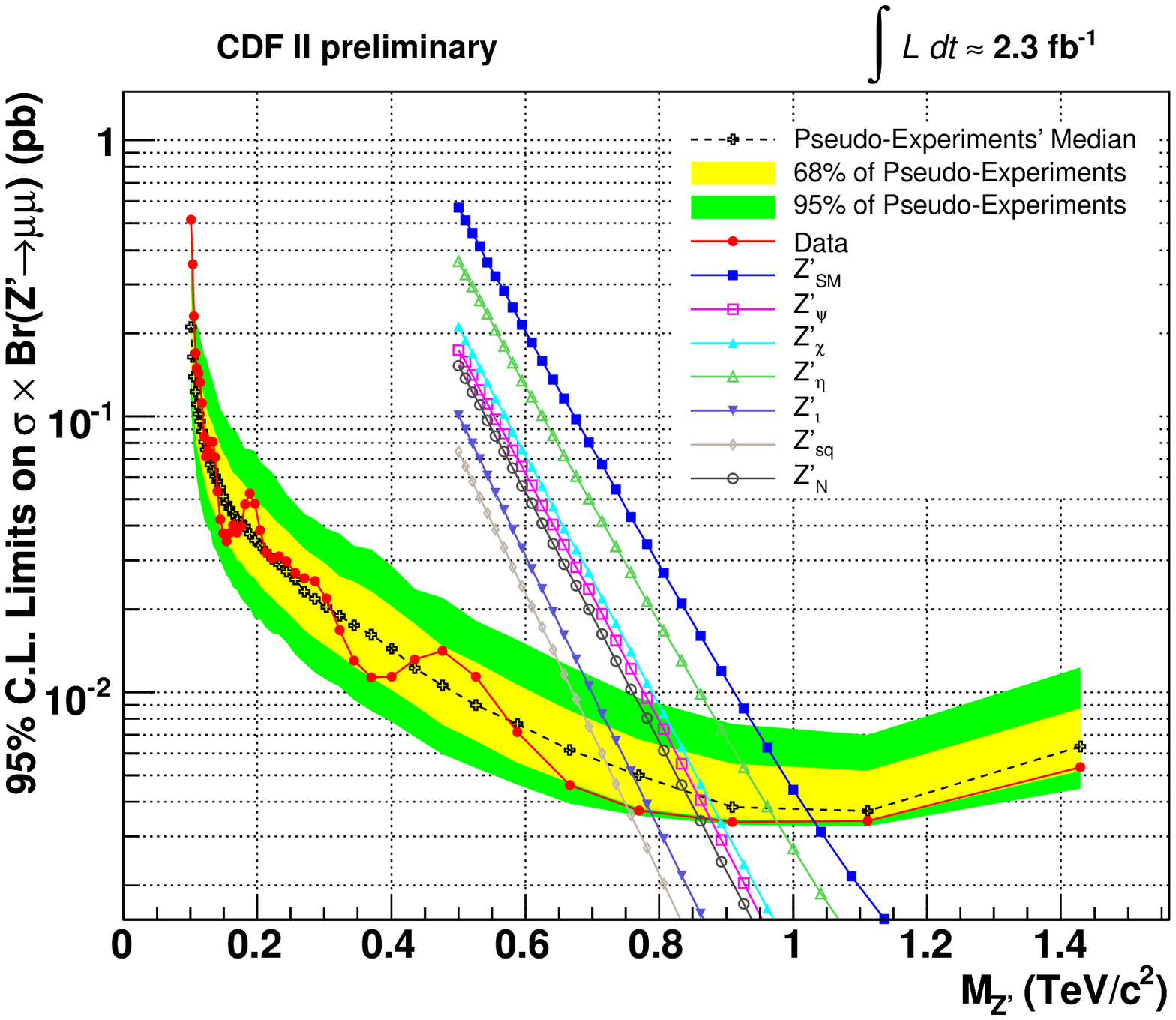}
  \includegraphics[width=0.49\hsize]
  {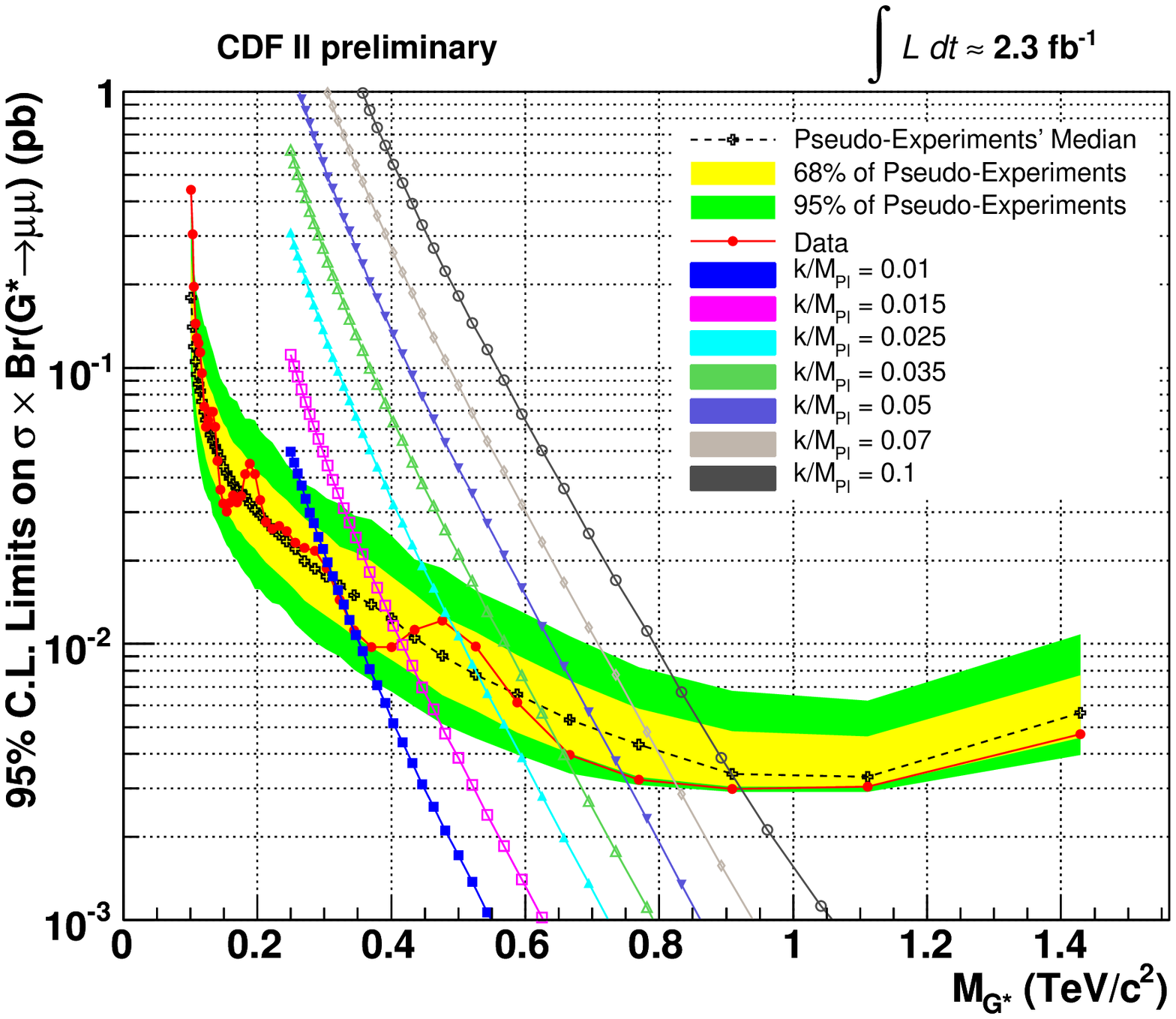}
  \end{tabular}
  \caption{The upper limits on $\sigma\cdot{\cal B}$ at 95\% C.L. 
    for spin-1 particles from the $m^+m^-$ data (left), 
    and on spin-2 particles from the $m^+m^-$ data (right).}
  \label{fig:mumu_limits}
\end{figure}

Fig.~\ref{fig:mumu} shows good agreement between data and expected backgrounds.
The 95\% C.L. limits on $\sigma\cdot{\cal B}(X\to \mu^+\mu^-)$
for spin-1 and spin-2 particles are shown in Fig.~\ref{fig:mumu_limits} together
with theoretical predictions for various $Z'$s and $G^*$s
with several $k/M_{pl}$ values.
This search provides the most stringent constraints on $Z'$s
in various models and on $G^*$.

% Our search strategy
% is to construct templates of the inverse invariant mass distribution
% for a range of Z' boson pole masses, add the background distributions
% to the templates, and compare the templates to the $1/m_{\mu\mu}$ distribution
% from data in the search region 1/m < 10 1/TeV. The simulated
% templates (including backgrounds) are normalized to the data in the 
% $70 < m_{\mu\mu} < 100$ GeV normalization region. We evaluate the binned
% Poisson likelihood as a function of signal template integral. The
% likelihood yields the best-fit number of signal events observed, as
% well as the confidence interval.

%%
%%
%%


\begin{thebibliography}{9}
\bibitem{rhot8_para} The following parameters are used: $M_8'=0$, $M(\pi_{22}^{8})=5M(\rho_{T8})/6$,
  $M(\pi_{22}^{1})=M(\pi_{22}^{8})/2$ and $M_8=5M(\rho_{T8})/6$.
\end{thebibliography}
\end{document}